\def\ln{\ell{n}}
\documentstyle[12pt]{article}
\begin{document}
\begin{titlepage} \vspace{0.2in} \begin{flushright}
MITH-97/05 \\ \end{flushright} \vspace*{1.5cm}
\begin{center} {\LARGE \bf  Mass generation of the lepton sector\\} 
\vspace*{0.8cm}
{\bf She-Sheng Xue$^{\dagger}$\\}
INFN, Section of Milan, Via Celoria 16, Milan, Italy\\
Physics Department, University of Milan, Italy\\
\vspace*{1.5cm}
{\bf   Abstract  \\ } \end{center} \indent
 
We discuss the necessity of a right-handed Weyl neutrino due to the vector-like
phenomenon of the regularized Standard Model. It is shown that this
right-handed neutrino is decoupled from low energies as a free particle, and
Dirac neutrino masses are very small. We suggest gauge-invariant couplings
between down quarks and charged leptons so that charged leptons acquire masses
without extra Goltstone modes. By examining Schwinger-Dyson equations for
lepton self-energy functions, we show that the neutrinos get their Dirac masses
via explicit symmetry breakings that attribute to the mixing between neutrinos
and charged leptons. An analysis of these Dyson equations gives the four
relationships between inter-generation mixing angles and lepton masses. 

\vfill \begin{flushleft} 1st March, 1997 \\
PACS 11.15Ha, 11.30.Rd, 11.30.Qc  \vspace*{1.5cm} \\
\noindent{\rule[-.3cm]{5cm}{.02cm}} \\
\vspace*{0.2cm} \hspace*{0.5cm} ${}^{\dagger}$ 
E-mail address: xue@milano.infn.it\end{flushleft} \end{titlepage}

\section{Introduction}
Since their appearance neutrinos have always been extremely peculiar. In the
sixty years of their life, their charge neutrality, their apparent
masslessness, their left-handedness have been at the centre of a conceptual
elaboration and an intensive experimental analysis that have played a major
role in donating to mankind the beauty of the electroweak theory. V-A theory
and Fermi universality would possibly have eluded us for a long time had the
eccentric properties of neutrinos, all tied to their apparent masslessness, not
captured the imagination of generations of experimentalists and theorists
alike. 

However, with the consolidation of the Standard Model (SM) and in particular
with the general views on (spontaneous?) mass generation in the SM, the
observed (almost) masslessness of the three neutrinos
($\nu_e,\nu_\mu,\nu_\tau$) has recently come to be viewed as a very problematic
and bizarre feature of the mechanism(s) that must be at work to produce the
very rich mass spectrum of the fundamental fields of the SM. Indeed, in the
(somewhat worrying) proliferation of the Yukawa couplings of fermions to the
Higgs fields that characterizes the generally accepted SM, no natural reason
can be found why the charge-neutral neutrinos are the fundamental particles of
the lowest mass; for in the generally accepted minimal Higgs mechanism, the
actual values of the fermion masses are in direct relation with the strengths
of their couplings to the Higgs doublet, and it appears rather bizarre that
nature has chosen to create a very sophisticated mass pattern by the mere
fine-tuning of a large number of parameters. 

\section{The ``No-Go'' theorem and high-dimension operators}

For more than a decade and half, it has been known that the left-handed
neutrino fields in the SM cannot consistently be defined in a
lattice-regularized quantum field theory. This theoretical inconsistency was
asserted by the rigorously demonstrated ``no-go'' theorem of Nielson and
Ninomiya\cite{nn}. This theorem states that under certain conditions, there
must exist exactly equal numbers of the left-handed neutrinos and right-handed
neutrinos and both handed neutrinos couple to gauge fields with the same
strength in the low-energy limit of a lattice-regularized quantum field theory,
if one insists on preserving chiral gauge symmetries. As a result, the neutrino
spectrum and the $W^\pm$-gauge coupling are no longer chiral (left-handed), but
rather vector-like (equally left- and right-handed). This vector-like
phenomenon is the generic feature of lattice-regularized chiral gauge
theories. 
In the basis of the ABJ anomaly, it was shown \cite{nn} that the absence of the
left-handed neutrino in the lattice-regularized SM is not an artifact of
lattice-regularization itself. In fact, the ``No-Go'' theorem shows a very
generic feature concerning the inconsistency of regularizing chiral gauge
theories in the high-energy region. 

In the low-energy region, on the other hand, the successful Standard Model
exhibits its very peculiar parity-violating features of purely left-handed
gauge coupling of the $W^\pm$-boson and only the left-handed neutrinos. The
``no-go'' theorem seems to run into a paradox that the experimentally
successful Standard model is in fact theoretically inconsistent. We do not
consider this paradox to be an intrinsic problem of the Standards Model.
Instead, we regard that this inconsistency may imply what are the Nature's
possible choices for the SM at short distances\cite{xue94,ns92} 

Since one of the prerequisites of the ``no-go'' theorem requires the
lagrangian to be bilinear in fermionic fields, as that of the SM,
this theorem strongly implies that the possible extensions of the
standard model in short distances are high-dimension operators in terms of
fermionic fields,
\begin{equation}
L_{\rm effective}=L_{SM}+{\rm high\!-\!dimension\!-\!operators},
\label{eff}
\end{equation}
that we call the effective lagrangian at a certain physical cutoff $\Lambda$.
This is to meant that if high-dimension operators are supplemented into the
standard model at short distances, the ``no-go'' theorem and resultant paradox
can be evaded {\it in principle}. However, it is difficult to show how this
``no-go'' theorem and the paradox can be evaded {\it in practice}. The
reasons are the following: (i) there can be many high-dimension
operators allowed by the chiral gauge symmetries of the 
SM\footnote{ Some of these operators explicitly violate the global
symmetries (e.g.~the baryon number) that are anomalous in the standard
model\cite{ep,cxue}.} in the high-energy region (the cut off); (ii) it
is a non-perturbative effort to determine where is the ultra-violet
fixed point in the space of high-dimension operators (couplings), 
which high-dimension operators are relevant to meet
with the needs of the phenomenologically successful SM at the low-energy 
region. 

From the phenomenological point of view, one of relevant dimension-6
operators should be the four-fermion interaction for the $\bar tt$ condensate
model\cite{bar} of the third generation of the quark sector ($a,b$ are the 
color 
indices),
\begin{equation}
G\bar Q^{ai}_L(x)\cdot t^a_R(x)\bar t^b_R(x)\cdot Q_L^{bi}(x); \hskip0.3cm 
Q_L=(t, b)_L,
\label{tt}
\end{equation}
which undergoes the spontaneous breaking of the chiral gauge symmetries of the
SM to generate the top quark mass that is much heavier than other quarks. In
principle, other quarks could have the same interaction as (\ref{tt}) at the
cutoff, we gave an interpretation\cite{xue96} why the interaction (\ref{tt}) is the only
one relevant in low energies. 

Analogously, since the $\tau$-lepton is most heaviest in the 
lepton sector, we could have the following
four-fermion interaction of the third generation of the lepton sector,
\begin{equation}
G\bar\psi^i_L(x)\cdot \tau_R(x) \bar\tau_R(x)\cdot \psi^i_L(x);\hskip0.3cm
\psi^i_L=(\nu_\tau, \tau)_L.
\label{tau}
\end{equation}
If this operator undergoes the spontaneous symmetry breaking, there would be
extra Goldstone bosons beside those from the $\bar tt$-condensate model.  
This situation is phenomenologically unacceptable. We make the following
observations\cite{xue96}: (i) the four-fermion couplings $G$ are equal 
in eqs.(\ref{tt}) and (\ref{tau}) for the reason 
of some underlying unification; (ii) the coupling $G$ 
in eqs.(\ref{tt}) is enhanced by the color factor $N_c=3$; (iii) if one 
fine-tunes $G$ in eq.(\ref{tt}) around ${N_cG_c\Lambda^2\over 2\pi^2}=4+0^+$, 
the spontaneous symmetry
breaking takes place and the 
operator (\ref{tt}) is relevant\footnote{At this point, this operator receives
anomalous dimension $\gamma_m=2$.}, while the operator (\ref{tau}) 
is irrelevant since the lepton sector is colorless ($N_c=1$) and 
the effective four-fermion coupling is bellow the threshold of taking place
the spontaneous symmetry breaking. This seems
not to run into the problem of extra Goldstone modes. These discussions can
be generalized to similar dimension-6 interaction for other charged leptons.
The question is how $\tau$ and other charged leptons acquire their masses?

Based on the chiral gauge symmetries of the SM, we are not only allowed
to have the high-dimension operators (\ref{tt},\ref{tau}), but also the 
four-fermion interaction between the quark and lepton sector,
\begin{equation}
G\bar\psi^i_L(x)\cdot \tau_R(x)\bar b^a_R(x)\cdot Q^{ai}_L(x).
\label{tau1}
\end{equation}
This operator, although it should be irrelevant in low energies, 
is clearly responsible for the $\tau$-lepton mass for any value of $G$, 
once the bottom quark is massive\footnote{It is discussed in
ref.\cite{xuem} that quark (except top) masses are generated by explicit 
symmetry breakings.}. 
We can have gauge invariant operators similar to eq.(\ref{tau1})for the first 
and second families as well. Tuning the four-fermion coupling to the critical
coupling $G\rightarrow G_c+0^+$, we obtain, at the cutoff,
\begin{eqnarray}
m_b &=& m_\tau\nonumber\\
m_s &=& m_\mu\nonumber\\
m_d &=& m_e.
\label{su5}
\end{eqnarray}
These are reminiscent of the predictions in the 
$SU(5)$ unification theory. 

Obviously, we give no explanations, from theoretical point of view, why the
operators should only be eqs.(\ref{tt},\ref{tau},\ref{tau1}) of the 
third
generation. One may conceives that the fermionic flavour symmetries of the
effective lagrangian in the high-energy region should be exact because of the
underlying physics that are quite possibly flavour blind, e.g.~quantum
gravity. For this reason, other possible dimension-6 operators comprising all
fermionic flavours in various generations cannot be certainly precluded 
from the effective lagrangian, as far as the chiral gauge symmetries of the SM
and naive dimensional counting are concerned. While, on the
other hand, due to the fact that the fermionic flavour symmetries of the
standard model is violently broken, and the relevant high-dimension operators
upon the ultra-violet fixed point of the effective lagrangian are presumably not
flavour symmetric. This circumstance is clearly governed by the properties and
complexities of the flavour dynamics of the effective lagrangian and 
its ground states\footnote{As for this
point, some discussions are presented in refs.\cite{xuem}}. 

In the context of the SM, the most crucial observation is that the
fermionic flavour symmetries should be broken explicitly rather than
spontaneously, otherwise we would have extra Goldstone modes, which are not
observed and not energetically favourable in the ground state. Thus, we
stipulate that the ultra-violet fixed point is such that other dimension-6
high-dimension operators except eq.(\ref{tt}), which develops the spontaneous
symmetry breaking, are irrelevant in the
low-energy limit. 

Even though the ``no-go'' theorem does not tell us what are the relevant
high-dimension operators of the SM in the low-energy limit, 
it really suggests us an existence of the right-handed neutrino
and necessary high-dimension operators at short distances without going into
the details of a concrete unification model. 

\section{On the smallness of neutrino masses}

As discussed, the right-handed neutrino $\nu_R$ is theoretically forced to
exist in the SM by the ``no-go'' theorem. However, it is
experimentally illusive and its couplings to the left-handed neutrinos, namely
Dirac neutrino masses, are very small contrasted sharply with other fermion masses. 
These two points implies us that the right-handed Weyl neutrinos $\nu_R$ should be
almost free particles, weakly couple to the left-handed neutrinos and other
particles. This means that each external right-handed
neutrino line of all interacting (1PI, one particle irreducible) operators should
be significantly suppressed in the low-energy limit. For this observation, we
stipulate that in all high-dimension
operators of the effective lagrangian (\ref{eff}), the right-handed neutrino 
fields appear as a ``high-dimension'' field defined as, 
\begin{equation}
\Delta\nu_R(x)\equiv\sum_\mu
\left[ \nu_R(x+\mu)+\nu_R(x-\mu)-2\nu_R(x)\right],
\label{diff}
\end{equation}
where the operator $``\Delta''$ is written as a discrete differentiation, 
and actually is a Dalambert's operator. Thus, the effective lagrangian 
(\ref{eff}) exactly is invariant under the transformations,
\begin{equation}
\bar\nu_R(x) \rightarrow \bar\nu_R(x)+\bar\epsilon,\hskip1.5cm
\nu_R(x) \rightarrow \nu_R(x)+\epsilon,
\label{shift}
\end{equation}
where $\epsilon$ is independent of space-time. As a result, in the effective 
lagrangian (\ref{eff}) there not exists dimension-6 operators involving
$\nu_R$ analogous to
(\ref{tt},\ref{tau}) for the $\tau$-lepton. 

In the most simplest case, we introduce only a single right-handed neutrino
$\nu_R$ that is a singlet of the chiral gauge symmetries of the SM\footnote
{It can be possible that each generation has its own right-handed neutrino.}. 
We discuss this problem in the third lepton generation, and it can be easily
generated into the first and second lepton generations
 
In the basis of similar arguments for eq.(\ref{tau1}) in the previous section,
except its kinetic term, this right-handed neutrino $\nu_R$ couples to 
the left-handed neutrino 
through the gauge invariant dimension-8 operator given by
\begin{equation}
G\bar\psi^i_L(x)\cdot\left[\Delta \nu_R(x)\right]
\bar t^a_R(x)\cdot Q_L^{ai}(x),
\label{neu}
\end{equation}
which could be responsible for neutrino's Dirac
masses. There are other possible high-dimension (at least dimension-10) gauge
invariant operators, which are relevant 
for doublers residing at the cut-off. The details of discussions presented
in ref.\cite{xue97} are out of the scope of this paper.

We turn to discuss the peculiar properties that Dirac neutrino masses are very small, 
the right-handed neutrino $\nu_R$ is a almost free
particle and decouple from all physical particles in the low-energy limit.
These properties can
be demonstrated by Ward identities of the $\nu_R$-shift-symmetry 
(\ref{shift}). The Ward
identity in terms of the primed fields corresponding to the 
$\nu_R$-shift-symmetry of the action (\ref{shift}) is given as 
\begin{equation}
\gamma_\mu\partial^\mu\nu'_R(x)
+G \langle\Delta\left(\bar Q^{ai}_L(x)\cdot
t^a_R(x)\psi_L^i(x)\right)\rangle-{\delta\Gamma\over\delta\bar
\nu'_R(x)}=0,
\label{w}
\end{equation}
where $``\Gamma''$ is the effective potential with non-vanishing external
sources; the prime field $\nu_R'\equiv\langle\nu_R\rangle$, where
$\langle\cdot\cdot\cdot\rangle$ is the expectation value respect to generating
function $Z(J,\eta)$. 
Based on this Ward identity, one can get all one-particle irreducible
(1PI) vertices containing at least one external $\nu_R$.

As the first example, one can
obtain an identity for the self-energy function $\Sigma^i(p)$, which is Dirac
mass for the $\tau$-neutrino. Performing a functional derivative of
eq.~(\ref{w}) with respect to the prime field $\psi'^i_L(0)$ and then putting
external sources $\eta=0$ and $J=0$, and we obtain
\begin{equation}
G\langle\Delta\left(\bar Q^{ai}_L(x)\cdot
t^a_R(x)\delta(x)\right)\rangle_\circ-
{\delta^2\Gamma\over\delta\psi'^i_L(0)
\delta\bar\nu'_R(x)}=0,
\label{ws1}
\end{equation}
where $\langle\cdot\cdot\cdot\rangle_\circ$ is the expectation value
with vanishing external sources $\eta$ and $J_\mu$.
Transforming into momentum space, we obtain
\begin{equation}
{1\over2}\Sigma^i(p)=2Gw(p)\langle\bar Q^{ai}_L(0)\cdot
t^a_R(0)\rangle_\circ=4w(p)m_t,\hskip0.3cm i={1\over2}
\label{ws2}
\end{equation}
where the well-known Wilson factor and the top-quark mass are,
\begin{eqnarray}
w(p)&\equiv&{1\over2}\int d^4xe^{-ipx}\Delta(x)
=\sum_\mu\left(1-\cos(p_\mu a)\right),\hskip0.2cm {\pi\over a}={\rm cutoff}
\nonumber\\
m_t&=&{G\over2}\langle\bar Q^{ai}_L(0)\cdot t^a_R(0)\rangle_\circ,
\label{wif}
\end{eqnarray}
where $m_t$ is the top-quark mass of the $\bar tt$-condensate model (\ref{tt}).
This clearly shows that in the low-energy limit, the self-energy function of 
Dirac $\tau$-neutrinos vanishes at the order of
\begin{equation}
\Sigma(p)\rightarrow O\left(({m_t\over\Lambda})^2m_t\right),\hskip0.3cm
 p\rightarrow m_t.
\label{small}
\end{equation}
This could be one of possible reasons for the smallness of Dirac neutrino
masses. 

As the second example, taking the functional derivative of eq.~(\ref{w}) with respect to $\nu'_R(0)$
and then putting external sources $\eta=0$ and $J=0$, we derive
\begin{equation}
(\gamma_\mu P_R)^{\beta\alpha}\partial^\mu \delta(x)
-{\delta^2\Gamma\over\delta\nu'^\alpha_R(0)
\delta\bar\nu'^\beta_R(x)}=0.
\label{wf}
\end{equation}
Thus, the two-point function in eq.~(\ref{wf}) is given as,
\begin{equation}
\int_xe^{-ipx}
{\delta^{(2)}\Gamma\over\delta\psi'_R(x)\delta\bar\psi'_R(0)}=i
\gamma_\mu p^\mu,\label{free}
\end{equation}
indicating that $\nu_R$ does not receive wave-function renormalization $Z_3$.

The third example is of the four-fermion interaction vertex. Analogously, 
one takes
functional derivatives of the Ward identity
(\ref{w}) with respect to $\psi'^i_L(0)$, $\bar Q'^{ai}_L(y)$ and $t'^a_R(z)$
and obtains 4-points interacting vertex involving an external $\nu_R'$,
\begin{equation}
\int_{xyz}e^{-iyq-ixp-izp'}
{\delta^{(4)}\Gamma\over\delta\psi'^i_L(0)\delta\bar Q'^{ai}_L(y)\delta t'^a_R(z)
\delta\bar\nu'_R(x)}=2Gw(p+{q\over 2}),
\label{4p}
\end{equation}
where $p+{q\over 2}$ are the momenta of the $\nu_R(x)$ field and $p'+{q\over 2}$ are
the momenta of the $t^a_R(x)$ field;
$p-{q\over 2}$ and $p'-{q\over 2}$ are the momenta of $\psi^i_L(x)$ field 
and $Q_L^{ai}(x)$ ($q$ is the momentum transfer.).
This interacting vertex vanishes in
the low-energy limit ($p,q\rightarrow 0$) for the same reason (\ref{small}). 
Further, as the consequence of
the Ward identity (\ref{w}), all 1PI n-point vertices
($n>4$) containing $\nu_R'$'s are just identical to zero.
\begin{equation}
{\delta^{(n)}\Gamma\over\delta^{(n-1)}(\cdot\cdot\cdot)\delta\bar\nu'_R(x)}
=0,\hskip0.3cm n>4.\label{n}
\end{equation}
where $\delta^{(n-1)}(\cdot\cdot\cdot)$ indicates $(n-1)$ derivatives with
respect to other prime (external) fields. 

These four identities eqs.(\ref{small},\ref{free},\ref{4p}) and (\ref{n})
show us two conclusions owing to the $\nu_R$-shift-symmetry:
\begin{enumerate}
\begin{itemize}
\item
the Dirac neutrino masses due to high-dimensions operators 
are extremely small;
\item 
the right-handed neutrino $\nu_R(x)$ in low-energies is a free particle
and decouples from other physical particles. 
\end{itemize}
\end{enumerate} 
These conclusions do not change if the gauge interactions of the SM 
are taken into account, since the right-handed neutrino introduced is a 
gauge singlet and the $\nu_R$-shift-symmetry must not be violated by
gauge interactions.

\section{Composite vector-like phenomenon}

In the effective lagrangian (\ref{eff}), the high-dimension operators implied
by the ``no-go''  theorem should in principle be all possible operators allowed
by the chiral gauge symmetries of standard model and the $\nu_R$-shift-symmetry
(\ref{shift}). In section 2 and 3, we only discussed the dimension-6 and
dimension-8 operators as far as the mass generation of quark and lepton sectors
is concerned. Beside, there must be operators whose dimension are larger than 8.
It is shown in ref.\cite{xue97} that we need the dimension-10 
operators to gauge-invariantly decouple unwanted ``doublers'' and
avoid the vector-like phenomenon in low energies. This means that some
dimension-10 operators should be relevant for ``doublers'' in the low-energy
limit. To be more specific,
the effective couplings of these dimension-10 operators certainly 
are momentum dependent.
When these effective couplings are larger than a certain threshold $\epsilon$ in
high-energies, three-fermion Weyl states with appropriate chiral quantum
numbers are bound\cite{ep,xue97}. These composite Weyl fermions couple to 
elementary Weyl
fermions to form gauge-invariantly massive Dirac fermions and all 1PI vertices
are vector-like consistently with the chiral gauge symmetries of the SM.
We call this scenario composite vector-like phenomenon. 

We will not enter
into the details of this issue to show all vector-like vertices and composite
spectra in
the high-energy region. Instead, inspired by this composite
vector-like phenomenon due to high-dimension operators in the high energy region, we
postulate an extension (model) of the standard model
beyond a certain energy scale $\epsilon$, 

\begin{enumerate}
\begin{itemize}

\item  right-handed three-fermion Weyl states possessing 
definite chiral gauge quantum
numbers of the $SU_L(2)\otimes U_Y(1)$ group are bound, and the threshold
$\epsilon$ associating with the binding energy is larger than the weak scale
$\Lambda_w(\sim 250$GeV);

\item  for given a conserved quantum number of the $SU_L(2)\otimes U_Y(1)$
gauge symmetries, the number of these composite Weyl states is equal to
the number of the elementary Weyl states, and they couple together to form
massive Dirac fermions;

\item the spectra and vertices are vector-like consistently with 
the $SU_L(2)\otimes U_Y(1)$ symmetries, and the $W^\pm$ gauge bosons 
possesses vector-like coupling to these composite Dirac fermions.

\end{itemize}
\end{enumerate}

Within the context of the third lepton generation, we explicitly discuss these three
assumptions. It is assumed there is an 
intermediate
energy-threshold $\epsilon$ between the cutoff and the weak scale ($v\sim 
250$GeV) of the spontaneous symmetry breaking, 
\begin{equation}
250GeV <\epsilon < \Lambda,
\label{epsilon}
\end{equation}
above this energy-threshold, the effective high-dimension operators 
are strong 
enough to form the three-fermion bound states that 
are given by\footnote{We do not discuss baryon number violating case,
where three-fermion states are anti-proton and anti-neutron\cite{cxue}.}
\begin{equation}
\nu^3_R\sim (\bar \nu_R\cdot \nu_L)\nu_R,\hskip0.3cm \tau^3_R\sim 
(\bar \tau_R\cdot \tau_L)\tau_R,
\label{threeb}
\end{equation}
which are right-handed Weyl fermions with the appropriate gauge quantum number
of the $SU_L(2)\otimes U_Y(1)$ symmetries. Whereas, to coincide with the 
parity-violating gauge coupling
observed in low-energies, at the threshold $\epsilon$ (\ref{epsilon}),
these three-fermion bound
states turn to three-fermion cuts, where they dissolve to their constituents 
(\ref{epsilon}) because of vanishing their binding energy. This intermediated
scale $\epsilon$ should be determined by effective high-dimension operators
(couplings) and vanishing the binding energy of 
three-fermion states. 

These three-fermion Weyl states (\ref{threeb}) couple to the elementary Weyl fields 
$\nu_L,\tau_L$ to form gauge invariant massive Dirac fermions,
\begin{equation}
\{\nu_L,\nu_R^3\}; \hskip1cm\{\tau_L,\tau^3_R\}.
\label{dirac}
\end{equation}
These massive Dirac fermions carry appropriate quantum numbers of
the $SU_L(2)$ symmetry and couple to the $W^\pm$ boson. The chiral
gauge symmetries of the SM is exact in high-energies, if we do not
consider the soft spontaneous symmetry breaking of the Higgs mechanism.

The above discussions are straightforwardly generalized to the first and
second generations.
This scenario of the composite vector-like phenomenon above the 
intermediate scale
(\ref{epsilon}) is reminiscent of the ``left-right'' symmetric extensions 
($SU_L(2)\otimes SU_R(2)\otimes U_{B-L}(1)$) of the
Standard model\cite{lr}. However, comparing with the ``left-right''
symmetric model, it should be noted that in this model, (i) the gauge 
symmetries are still $SU_L(2)\otimes U_Y(1)$; (ii) there are no needs of 
new elementary fermions and gauge bosons accommodated by the $SU_R(2)$ gauge group;
(iii) the intermediate scale $\epsilon$ is not due to spontaneous symmetry
breakings, no Goldstone bosons associate with the form of three-fermion
states at the scale $\epsilon$. 

In this early stage, we make no attempt to give a complete
description of various effective vertices (1PI) in this model.  
In this section, we wish to reconsider the
self-energy functions (Dirac masses) of neutrinos
$\Sigma_{\nu_i}(p)$ and charged leptons $\Sigma_{l_i}(p)$ by taking into
account the possible relevant 1PI vertices function raised in the high-energy
region of this model. 
The right-handed fermion states are the mixing states comprising
the elementary state $\nu_R$ ($\tau_R$) and the composite state $\nu^3_R$ 
($\tau^3_R$): 
\begin{equation}
\Psi^\nu_R=(\nu_R,\nu^3_R);\hskip0.5cm \Psi^\tau_R=(\tau_R,\tau^3_R).
\label{mixing}
\end{equation}
The composite Dirac particle instead of (\ref{dirac}) are then given,
\begin{equation}
\Psi^\nu_D=\{\nu_L,\Psi^\nu_R\};\hskip0.5cm \Psi^\tau_D=\{\tau_L,\Psi^\tau_R\}.
\label{mixingd}
\end{equation}
If the soft spontaneous breaking of chiral gauge symmetries is introduced,
the self-energy functions $\Sigma_\nu(p)$ 
($\Sigma_l(p)$) are coupling between $\nu_L(\tau_L)$ and mixing 
right-handed fermion states $\Psi^\nu_R(\Psi^\tau_R)$ given by (\ref{mixing}).
This clearly modifies the self-energy functions $\Sigma_\nu(p)$ 
($\Sigma_l(p)$) of neutrinos and
charged leptons, which were the couplings (mass operators) between $\nu_L(\tau_L)$ and 
$\nu_R(\tau_R)$ in the SM.

According to eq.(\ref{mixingd}), the modified self-energy functions are the
effective vertices coupling between the $\nu_L(\tau_L)$ and mixing
states $\psi^\nu_R(\psi^\tau_R)$. The effective gauge coupling of 
$W^\pm$-bosons to composite Dirac fermions is vector-like in the high-energy 
region. Together with $W^\pm$'s purely left-handed
gauge coupling observed in the low-energy region, one can write an effective gauge
coupling as, 
\begin{eqnarray}
\Gamma^{ij}_\mu (q)&=&i{g_2\over2\sqrt{2}}V_{ij}\gamma_\mu (P_L+f(q))\label{wv}\\
f(q)&\not=&0,\hskip0.5cm q\ge\epsilon,
\end{eqnarray}
where $g_2$ is the $SU_L(2)$ coupling.
In eq.(\ref{wv}), the non-vanishing of the vector-like vertex function
$f(q)$ in the high-energy region $\epsilon <q<\Lambda$ is clearly related to the 
existence
of the three-fermion states (\ref{threeb}). 
Upon the energy threshold (\ref{epsilon}) where the three-fermion states 
turn to three-fermion cuts and dissolve into their constituents,
the effective vertex function $f(q)$ must vanishes,
\begin{equation}
f(q)|_{q\rightarrow\epsilon+0^+}\rightarrow 0.
\label{threshold}
\end{equation}
The $V_{ij}$ in eq.(\ref{wv}) is the CKM-matrix\cite{ckm}, since all 
fermionic states 
discussed are not the eigenstates of the chiral gauge symmetries of the SM.

Because of the effective gauge coupling (\ref{wv}), we find that the 
$W^\pm$ bosons have 
the contributions
to the Schwinger-Dyson equations for the self-energy functions 
$\Sigma_\nu(p)$ and $\Sigma_l(p)$ for $(p\ge\epsilon)$. 
We can approximately write the $W$-boson's contributions:
\begin{eqnarray}
W_{\nu_i}(p)&=&\left({g_2\over2\sqrt{2}}\right)^2
|V_{ij}|^2\int_{|p'|\ge\epsilon} {f(p'-p)\over (p-p')^2+M_w^2}
{\Sigma_{l_j}(p'^2)\over p'^2+\Sigma_{l_j}^2(p'^2)},\nonumber\\
W_{l_j}(p)&=&\left({g_2\over2\sqrt{2}}\right)^2
|V_{ji}|^2\int_{|p'|\ge\epsilon} {f(p'-p)\over (p-p')^2+M_w^2}
{\Sigma_{\nu_i}(p'^2)\over p'^2+\Sigma_{\nu_i}^2(p'^2)},
\label{ww}
\end{eqnarray}
where the integration of the internal momentum $p'$
starts from the intermediate threshold $\epsilon$ to the cut-off $\Lambda$.

\section{Mass generation of the lepton sector}

With the $W$-boson's contributions (\ref{ww}), 
the Schwinger-Dyson equations for the self-energy functions of
the neutrinos and charged leptons
respectively turn out to be highly non-trivial and coupled,
\begin{eqnarray}
\Sigma_{\nu_i}(p)&=&W_{\nu_i}(p);\label{selft}\\
\Sigma_{l_j}(p^2)&=&\Sigma_{q_j}(\Lambda)
+W_{l_j}(p)+3e^2\int^\Lambda_{p'} {1\over (p-p')^2}
{\Sigma_{l_j}(p')\over p'^2+\Sigma_{l_j}^2(p'^2)},\label{selfb}
\end{eqnarray}
where the bare down quark masses are given as,
\begin{equation}
\Sigma_{q_j}(\Lambda)=m_d(\Lambda), m_s(\Lambda), m_b(\Lambda),
\label{quark}
\end{equation}
are due to the four-fermion interaction
(\ref{tau1}) and the contribution of (\ref{neu}) to neutrinos masses is 
neglected.
This $W(p)$'s contributions are perturbative additions to the original
Schwinger-Dyson equations of the SM. 
One can see that eq.(\ref{ww}) mixes up
the Schwinger-Dyson equations for fermionic self-energy functions of different
generations and charge sectors. We have no reason to put the CKM-matrix
$V_{ij}=\delta_{ij}$, since the mixing between generations could be very large.

To solve the integral equations (\ref{selft},\ref{selfb}), one way is to
divide them into two integral equations corresponding to the regions
$p\in (0,\epsilon)$ and $p\in (\epsilon, \Lambda)$ respectively, and use
the continuation of self-energy functions $\Sigma(p)$ at the scale 
$\epsilon$ to match two solutions. Here, we alternatively adopt a
simple and approximate way to solve these coupled integral equations.
Assuming the scale $\epsilon$ is large enough and $p'>\epsilon\gg 1$, 
we approximate eqs.(\ref{ww})
to be,
\begin{equation}
W_{\nu_i}(p)\simeq \alpha_w(p)|V_{ij}|^2\Sigma_{l_j}(\Lambda),\hskip0.5cm
W_{l_j}(p)\simeq \alpha_w(p)|V_{ji}|^2\Sigma_{\nu_i}(\Lambda),
\label{www}
\end{equation}
where
\begin{equation}
\alpha_w(p)\simeq\left({g_2\over2\sqrt{2}}\right)^2
\int^\Lambda_{|p'|\ge\epsilon} {f(p'-p)\over (p-p')^2+M_w^2}
{1\over p'^2}.
\label{alpha}
\end{equation}
For the low-energy $p\ll\epsilon$, assuming $f(p')\simeq f$, as a 
slow-varying (small) function of $p'$($p'>\epsilon$), we get 
\begin{equation}
\alpha(p)={\alpha_2\over16\pi}f\ln{\Lambda\over\epsilon},\hskip0.2cm 
\alpha_2={g_2^2\over4\pi}.
\label{alpha2}
\end{equation}
For the high-energy $p>\epsilon\gg 1$, we approximately set
\begin{equation}
\alpha_w(p)\simeq\alpha_w(\Lambda),
\label{al}
\end{equation}
as an unknown constant. 

Using eqs.(\ref{selft},\ref{alpha2}), we obtain the three relations 
(gap-equations) between neutrino masses and charged lepton masses,
\begin{equation}
\Sigma_{\nu_i}(p)={\alpha_2\over16\pi}f\ln{\Lambda\over\epsilon}
|V_{ij}|^2\Sigma_{l_j}(\Lambda),\hskip0.3cm p\ll\epsilon.
\label{neutri}
\end{equation}
We find that neutrino masses are related to charged lepton masses via 
flavour mixing, and neutrino masses are zero, if the intermediated scale 
$\epsilon=\Lambda$, which means no vector-like phenomenon described in the
previous section. 

We can straightforwardly solve the coupled integral equation (\ref{selfb}) of
charged leptons in the high-energy region. In the ultraviolet region ($x=p^2\gg
1$), the nonlinearity is negligible and the integral eq.(\ref{selfb}) can be
converted to the following boundary value problem\cite{kogut} 
\begin{eqnarray}
{d\over dx}\left(x^2\Sigma'_{l_j}(x)\right)+{\alpha\over 4\alpha_c}
\Sigma_{l_j}(x)&=&0,
\label{deqb'}\\
\Lambda^2\Sigma'_{l_j}(\Lambda^2)+\Sigma_{l_j}(\Lambda^2)&=&
\Sigma_{q_j}(\Lambda)+\alpha_w(\Lambda)|V_{ji}|^2\Sigma_{\nu_i}(\Lambda),
\label{boundaryb'}
\end{eqnarray}
These are differential equations with the coupled inhomogeneous
boundary conditions at the cutoff. Those inhomogeneous terms act as
bare mass terms in the integral equation (\ref{selfb}).

The generic solution to eq.(\ref{deqb'}) for $(x\gg 1)$ 
is given\cite{kogut}
\begin{equation}
\Sigma_{l_j}(x) \simeq {A_{l_j}\mu^2\over\sqrt{ x}}{\rm sinh}
\left({1\over2}\sqrt{
1-{\alpha\over\alpha_c}}\ln({x\over\mu^2})\right),
\label{solution}
\end{equation}
where $A_{l_j}$ are arbitrary constants, and $\mu$ is an inferred scale.
Thus, we obtain the gap-equation of this coupled system from the boundary
condition (\ref{boundaryb'}),
\begin{equation}
\alpha_w(\Lambda)|V_{ji}|^2\Sigma_{\nu_i}(\Lambda)=
{A_{l_j}\mu^2\over2\Lambda}\left[{\rm sinh}\theta+
\sqrt{1-{\alpha\over\alpha_c}} {\rm cosh}\theta\right]-\Sigma_{q_j}(\Lambda),
\label{boundary2f}
\end{equation}
where 
\begin{equation}
\theta={1\over2}\sqrt{
1-{\alpha\over\alpha_c}}\ln({\Lambda^2\over\mu^2}),
\label{theta}
\end{equation}
The first conclusion can be derived from these gap-equations 
(\ref{neutri},\ref{boundary2f}) is that
if the down quarks are massive, 
the self-energy functions of the neutrinos, charged leptons must be non-trivial 
\begin{equation}
\Sigma_{\nu_i}(\Lambda)\not=0;\hskip0.3cm {\rm and } \hskip0.3cm 
\Sigma_{l_j}(\Lambda)\not=0,
\end{equation}
they are generated by the explicit symmetry breaking\cite{mn}.

We turn to find the solution of the gap-equation 
(\ref{boundary2f}) in the low-energy limit ($\mu\ll\Lambda$). 
Using eq.(\ref{solution}) for $\Sigma_{l_j}(\Lambda)$, we obtain 
the gap-equations for $\mu\ll\Lambda$:
\begin{equation}
\alpha_w(\Lambda)|V_{ji}|^2\Sigma_{\nu_i}(\Lambda)=\Sigma_{l_j}(\Lambda)
-{1\over4}{\alpha\over\alpha_c}\Sigma_{l_j}(\Lambda)
-\Sigma_{q_j}(\Lambda).
\label{low3b'}
\end{equation}
Since down quark and charged lepton masses at the cutoff are
equal (\ref{su5}) if gauge interactions are turned off, we have a cancellation 
in the RHS of gap-equation (\ref{low3b'}), as a result,
\begin{eqnarray}
\alpha_w(\Lambda)|V_{ij}|^2\Sigma_{l_j}(\Lambda)&=&
\Sigma_{\nu_i}(\Lambda);\label{low3''}\\
\alpha_w(\Lambda)|V_{ji}|^2\Sigma_{\nu_i}(\Lambda)&=&
-{1\over4}{\alpha\over\alpha_c}\Sigma_{l_j}(\Lambda).
\label{low3b''}
\end{eqnarray}
These are three gap-equations relating neutrino and charged lepton masses at 
the cutoff.

\section{Lepton masses and mixing angles}

In previous section, we obtained six gap-equations (\ref{neutri},\ref{low3b''})
relating neutrino and charged lepton masses at the cutoff. Noticing the mass
ratio of fermions in the same charge sector (but different generations) should
be scaling invariant (renormalization group invariant), we take ratios between
two equations of the gap-equations (\ref{neutri}), and two equations of the
gap-equations (\ref{low3b''}). We arrive at: 
\begin{eqnarray}
{m_{\nu_e}\over m_{\nu_\mu}}&=&{|V_{\nu_ee}|^2m_e+|V_{\nu_e\mu}|^2m_\mu
+|V_{\nu_e\tau}|^2m_\tau\over
|V_{{\nu_\mu}e}|^2m_e+|V_{{\nu_\mu}\mu}|^2m_\mu
+|V_{{\nu_\mu}\tau}|^2m_\tau},\label{r1}\\
{m_{\nu_e}\over m_{\nu_\tau}}&=&{|V_{\nu_ee}|^2m_e+|V_{\nu_e\mu}|^2m_\mu
+|V_{\nu_e\tau}|^2m_\tau\over
|V_{{\nu_\tau}e}|^2m_e+|V_{{\nu_\tau}\mu}|^2m_\mu
+|V_{{\nu_\tau}\tau}|^2m_\tau},\label{r2}\\
{m_{\nu_\mu}\over m_{\nu_\tau}}&=&{|V_{{\nu_\mu}e}|^2m_e
+|V_{{\nu_\mu}\mu}|^2m_\mu
+|V_{{\nu_\mu}\tau}|^2m_\tau\over
|V_{{\nu_\tau}e}|^2m_e+|V_{{\nu_\tau}\mu}|^2m_\mu
+|V_{{\nu_\tau}\tau}|^2m_\tau},\label{r1'}
\end{eqnarray}
and
\begin{eqnarray}
{m_e\over m_\mu}&=&{|V_{e\nu_e}|^2m_{\nu_e}+|V_{e{\nu_\mu}}|^2m_{\nu_\mu}
+|V_{e{\nu_\tau}}|^2m_{\nu_\tau}\over
|V_{\mu\nu_e}|^2m_{\nu_e}+|V_{\mu{\nu_\mu}}|^2m_{\nu_\mu}
+|V_{\mu{\nu_\tau}}|^2m_{\nu_\tau}},\label{r3}\\
{m_e\over m_\tau}&=&{|V_{e\nu_e}|^2m_{\nu_e}+|V_{e{\nu_\mu}}|^2m_{\nu_\mu}
+|V_{e{\nu_\tau}}|^2m_{\nu_\tau}\over
|V_{\tau\nu_e}|^2m_{\nu_e}+|V_{\tau{\nu_\mu}}|^2m_{\nu_\mu}
+|V_{\tau{\nu_\tau}}|^2m_{\nu_\tau}},\label{r4}\\
{m_\mu\over m_\tau}&=&{|V_{\mu\nu_e}|^2m_{\nu_e}+|V_{\mu{\nu_\mu}}|^2m_{\nu_\mu}
+|V_{\mu{\nu_\tau}}|^2m_{\nu_\tau}\over
|V_{\tau\nu_e}|^2m_{\nu_e}+|V_{\tau{\nu_\mu}}|^2m_{\nu_\mu}
+|V_{\tau{\nu_\tau}}|^2m_{\nu_\tau}}.\label{r3'}
\end{eqnarray}
In these equations, all fermion masses are defined at the same low-energy
scale. There are only four independent equations that completely determine the
four CKM mixing angles in terms of six lepton masses. 
 
The analysis of these four equations to find explicit relations between masses
and mixing angles will be presented in the coming paper soon. This paper is 
written for the proceeding of the 1997 Shizuoka workshop on masses and 
mixings of quarks and leptons (March 19-21). I thank Prof.~Yoshio Koide and
other organizers for providing me the financial support to participate this
workshop.

\end{document}